\documentclass[preprint]{aa}
\usepackage{graphicx}
\usepackage{txfonts}

\begin{document}

\title{Optimum estimate of delays and dispersive effects in low-frequency interferometric observations}

\titlerunning{Optimum estimate of delays and dispersive effects}

\date{Submitted on 2009 Dec 23. Accepted on 2010 May 3.}
\author{I. Mart\'i-Vidal}
\institute{Max-Planck-Institut f\"ur Radiastronomie, Auf dem H\"ugel 69, D-53121 Bonn (Germany) 
          \email{imartiv@mpifr.de}}

\abstract
{Modern radio interferometers sensitive to low frequencies will make use of 
wide-band detectors (with bandwidths of the order of the observing frequency) and 
correlators with high data processing rates. It will be possible to simultaneously 
correlate data from many sub-bands spread through the whole bandwidth of the detectors. 
For such wide bandwidths, dispersive effects from the atmosphere introduce variations 
in the fringe delay which change through the whole band of the receivers. These undesired 
dispersive effects must be estimated and calibrated with the highest precision.}
{We studied the achievable precision in the estimate of the ionospheric dispersion and 
the dynamic range of the correlated fringes for different distributions of sub-bands in 
low-frequency and wide-band interferometric observations. Our study is focused on the 
case of sub-bands with a bandwidth much narrower than that of the total covered spectrum 
(case of LOFAR).}
{We computed the formal statistical uncertainty of the ionospheric delay, the delay 
ambiguity and the dynamic range of the correlated fringes using four different kinds of 
distributions of the sub-bands: constant spacing between sub-bands, random spacings, 
spacings based on a power-law distribution, and spacings based on Golomb rulers 
(sets of integers, $n_i$, whose sets of differences, $n_j-n_i$, have non-repeated elements).}
{We compare the formal uncertainties in the estimate of ionospheric effects in the 
data, the ambiguity of the delays, and the dynamic range of the correlated fringes
for the four different kinds of sub-band distributions.}
{For a large number of sub-bands ($> 20$, depending on the delay window) 
spacings based on Golomb rulers give the most precise estimates of dispersive effects and 
the highest dynamic ranges of the fringes. Spacings based on the power-law distribution give 
similar (but slightly worse) results, although the results are better than those from the 
Golomb rulers for a smaller numbers of sub-bands. Random distributions of sub-bands result in 
relatively large dynamic ranges of the fringes, but the estimate of dispersive effects through 
the band is worse. A constant 
spacing of sub-bands results in very bad dynamic ranges of the fringes, but the estimates 
of dispersive effects have a precision similar to that obtained with the power-law 
distribution. Combining all the results, the power-law distribution gives the best 
compromise between homogeneity in the sampling of the bandwidth, precision in the estimate 
of the ionospheric dispersive effects, dynamic range of the correlated fringes, and 
ambiguity of the group delay.}


 \maketitle

\section{Introduction}

Modern radio interferometers sensitive to low frequencies, like the LOw Frequency ARray
(LOFAR, see, e.g., de Vos, Gunst, \& Nijboer \cite{Vos2009} and references therein), will make use of 
wide-band detectors (with a bandwidth of the order of the 
observing frequency) and correlators with a high data processing rate. For the case of 
LOFAR, it will be possible to correlate a bandwidth of 42\,MHz, which can be divided 
into many sub-bands to be spread through a wide band (either 10--90\,MHz, 110--190\,MHz, 
170--230\,MHz, or 210--250\,MHz; de Vos, Gunst, \& Nijboer \cite{Vos2009}). Therefore, the 
correlator will be able to simultaneously cover a large 
portion of the low-frequency spectrum. However, atmospheric dispersion may introduce 
strong frequency-dependent effects, which will be so different for each sub-band and must 
be estimated and calibrated with care. At low frequencies, the ionosphere is the main 
limiting factor in the quality of the interferometric observations. For the calibration of 
the ionospheric dispersion, the Total Electron Content (TEC) must be estimated through the
field of view (FoV) over each antenna and for each time. The TEC can be estimated, for 
instance, from Global-Positioning-System (GPS) data (e.g. Ros et al. \cite{Ros2000};
Todorova, Hobiger, \& Schuh \cite{Todorova2006}). With these techniques, however, it is 
not possible to obtain high-resolution estimates of the ionospheric turbulent screen over 
the FoV. For that purpose, it is necessary to apply self-calibration strategies, using 
therefore the interferometric data to derive the structure of the TEC screen over 
each station and its evolution (see Intema et al. \cite{Intema} and Cohen \& Rottgering 
\cite{Cohen} for a discussion on the ionospheric screens).  

The distribution of sub-bands through the whole bandwidth of the detectors
affects the scientific information that can be extracted from the observations, but is
also important for a precise and accurate estimate of the atmospheric dispersion from the 
interferometric observables. The distribution of sub-bands has also an effect in delay 
space, since the interferometric fringes have a shape related to the Fourier transform of 
the bandpass (i.e., in our case, the distribution of sub-bands). Therefore, we should 
distribute the sub-bands in such a way that the spectral coverage and sampling are 
maximized, but the dynamic range of the fringes (i.e., the height of the fringe peak
relative to that of the highest sidelobe) is also maximized to improve the sensitivity of 
the interferometer. Finding out the right distribution of sub-bands to achieve an optimum 
spectral sampling and fringe dynamic range is not that simple, and the answer may depend
on each case, namely, the total bandwidth to be covered, the number of sub-bands, and 
the bandwidth of each sub-band. For instance, Petrachenko (\cite{Petrachenko2008}) studied 
the performance of ``broadband delays'', which are computed from several bands 
(up to 5) spread from 2$-$3\,GHz to 11$-$14\,GHz, as a function of the way these bands 
are distributed through the spectrum. Petrachenko (\cite{Petrachenko2008}) concluded 
that the use of more than 2 bands, covering a total bandwidth as wide as possible, improves 
the performance of the interferometer.

In this paper, we report on a study of the spectral coverage, the precision in the 
estimate of atmospheric dispersive effects, the delay ambiguity, the dynamic range of 
the fringes, and the precision in the estimate of the source spectral index for different 
kinds of sub-band distributions and spectral configurations of 
an interferometer at low frequencies. The remainder of this paper is structured as follows: 
in Sect. \ref{II} we describe the process of analysis. In Sect. \ref{III} we describe the 
different sub-band distributions studied. In Sect. \ref{IV} we report on the 
results obtained and in Sect. \ref{V} we summarize our conclusions.

\section{Analysis}
\label{II}

\subsection{Spectral configuration of the interferometer}

The spectral configuration of an interferometer can be characterized using
the following parameters: i) minimum observing frequency, $\nu_\mathrm{m}$, ii) 
total bandwidth in units of the minimum observing frequency, i.e., 

\begin{equation}
\beta = (\nu_\mathrm{M} - \nu_\mathrm{m})/\nu_\mathrm{m} ,
\label{BetaEq}
\end{equation}

\noindent where $\nu_\mathrm{M}$ is the maximum observing frequency, iii) 
number of sub-bands, $N$ (and their distribution), and iv) the bandwidth 
of each sub-band, $\Delta \nu$. The $i$th sub-band is centered at 
frequency $\nu_i$ (with $\nu_\mathrm{m} < \nu_i < \nu_\mathrm{M}$), which can
be written as

\begin{equation}
\nu_i = (1 + \beta\,R_i)\,\nu_\mathrm{m} ,
\end{equation}

\noindent being $R_i$ a real number between $0.5 \Delta \nu/(\beta \nu_\mathrm{m})$ 
and $1 - 0.5 \Delta \nu/(\beta \nu_\mathrm{m})$. The spectral 
configuration of an interferometric observation is then characterized
by $\nu_\mathrm{m}$, $\beta$, $\Delta \nu_i$, and $R_i$. For simplicity, we 
assume the same $\Delta \nu_i = \Delta \nu$ for all sub-bands, and we 
also assume that $\Delta \nu << \beta \nu_\mathrm{m}$ (i.e., the bandwidth
of the sub-bands is much narrower than the total covered bandwidth, so $R_i$ 
is defined between 0 and 1).
This latter assumption corresponds to the case of the LOFAR 
interferometer.

\subsection{Dynamic range of the fringes}

The response of a baseline of an interferometer to a set of observed sources
is equal to the addition of several fringes, one fringe for each source. The
amplitude of these fringes is equal to a shrinked version of the amplitude of the 
Fourier transform of the bandpass (e.g., Thomson, Moran \& Swenson \cite{TMS91}). 
Spectral effects in the sources are not considered here. In 
the case of several sub-bands spread through a wider band, different shapes can be obtained 
for the fringes, which may have relatively large sidelobes. These large sidelobes may 
lead to confusion in the estimate of the fringe peaks. In frequency space, this effect 
can be understood as the possibility of fitting different slopes of phase vs. 
frequency to the same dataset. Minimizing the height of the sidelobes and maximizing 
their distance to the fringe peak decreases the probability of confusion and, therefore, 
enhances the sensitivity of the interferometer. 

For each of the studied band distributions (described in Sect. \ref{III}), the shape of 
the fringes, $F$, was computed as a vector with elements given by

\begin{equation}
F_k = \left|\sum_{i=1}^{N}{\cos{\left( \pi \frac{\nu_i}{\beta \nu_\mathrm{m}}\,(k-1)\right)}}\right| ,
\end{equation}

\noindent i.e., the module of the Direct Fourier Transform (DFT) of the sub-band distribution
(the sine term of the DFT is zero). The vector $F$ is thus computed assuming that the bandwidth 
of the sub-bands, $\Delta \nu$ is much narrower than the total bandwidth, $\beta \nu_\mathrm{m}$. The 
dynamic range of the fringe is estimated as the ratio between the fringe peak (which corresponds
to the element $F_1$) and the peak of the highest sidelobe. We call this ratio $D$. The ambiguity 
of the delay is computed as the distance between these two peaks, in units of the Nyquist time 
resolution (i.e., $1/(2\beta\nu_\mathrm{m})$). We call this quantity $\tau_\mathrm{amb}$. For the results reported
in the following sections, we used a vector $F$ of 512 elements.

If the width of the sub-bands is not much narrower than the total covered bandwidth,
the dynamic range, $D$, is corrected by 

\begin{equation}
D \rightarrow D / \textrm{sinc}{\left( \pi \frac{\Delta \nu}{2\beta\nu_\mathrm{m}}\tau_\mathrm{amb} \right)} ,
\end{equation}

\noindent where $\textrm{sinc}(x) = \sin{(x)}/x$.

\subsection{Estimate of the atmospheric dispersion}

The ionosphere introduces a change in the phase of the visibilities. For a given baseline
and source, this phase depends on frequency as (e.g., Thomson, Moran \& Swenson \cite{TMS91}) 

\begin{equation}
\Delta \phi_i = K\,\nu_i^{-1} .
\label{IonPhi}
\end{equation}

The parameter $K$ is related to the TEC of the ionosphere over the 
elements of the baseline in the line-of-sight direction to the source. For a good calibration
of the ionospheric dispersion, $K$ must be precisely estimated for each baseline, time, and
pointing direction over the FoV of the interferometer.

The formal statistical uncertainty in the estimate of $K$ is that of the slope of the 
linear fit of $\phi_i$ vs. $\nu_i^{-1}$. It is straightforward to show that this uncertainty
is

\begin{equation}
\sigma(K) \propto ( <\nu_i^{-2}> - <\nu_i^{-1}>^2)^{-1} = \sigma(\nu_i^{-1})^{-1} ,
\label{SigK}
\end{equation}

\noindent i.e., the precision in $K$ is proportional to the standard deviation of the distribution
of the inverse of the central frequencies of the sub-bands, $\sigma(\nu_i^{-1})$. This standard deviation
maximizes when half of the sub-bands gather close to $\nu_\mathrm{m}$ and the other half gather close to 
$\nu_\mathrm{M}$.
In this case

\begin{equation}
\sigma(\nu_i^{-1})_\mathrm{max} = 0.5\,(\nu_\mathrm{m}^{-1} - \nu_\mathrm{M}^{-1}) .
\label{minSig}
\end{equation}

However, this distribution of sub-bands is not, by far, optimum, since the sampling of the total band 
is very poor and, moreover, spectral effects in the sources, which would not be well sampled through 
the bandwidth, could introduce important systematics in the estimate of the ionospheric dispersion 
using Eq. \ref{IonPhi}. Additionally, there could be several undetermined $2\pi$-cycles of the 
phase drift caused by the ionosphere between $\nu_\mathrm{m}$ and $\nu_\mathrm{M}$, so, for this distribution, it would 
not be possible to connect the phases between all sub-bands to obtain a correct estimate of $K$.

The uncertainties in the estimate of the ionospheric dispersion, which we analyze in Sect. \ref{IV}, 
were estimated as $\sigma(K)$ (computed from Eq. \ref{SigK}) in units of its minimum 
possible value, $\sigma(K)_\mathrm{min}$ (i.e., that one corresponding to $\sigma(\nu_i^{-1})_\mathrm{max}$, 
which is given in Eq. \ref{minSig}).

\section{Distributions of sub-bands}
\label{III}

\begin{figure*}
\centering
\includegraphics[width=16cm]{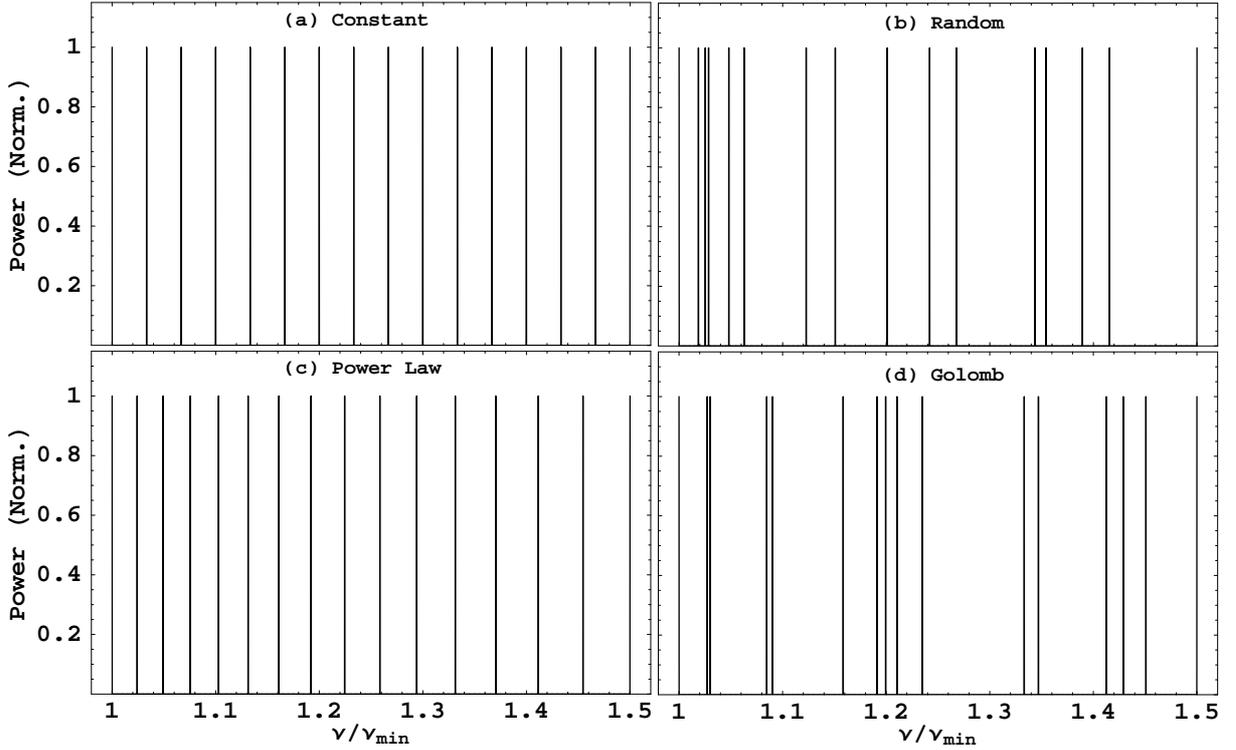}
\caption{Examples of the four kinds of distributions studied in this paper (using 16 sub-bands 
and setting $\beta = 0.5$, see Eq. \ref{BetaEq}). (a) corresponds to the uniform (i.e. constant) 
distribution, (b) to the random distribution, (c) to the power-law distribution, and (d) to the 
distribution based on the Golomb ruler.}
\label{DistriFig}
\end{figure*}

\subsection{Uniform (i.e., constant) distribution}

This is the most simple spectral configuration of the interferometer. The central frequencies of 
the sub-bands are distributed as 

\begin{equation}
\nu_i = \left(1 + \beta\frac{i-1}{N-1} \right)\,\nu_\mathrm{m} ,
\label{HomogEq}
\end{equation}

\noindent i.e., the frequency spacing between sub-bands is constant. We show an example of this 
distribution in Fig. \ref{DistriFig}(a).

\subsection{Random distribution}

In this case, the sub-bands are randomly distributed over the bandwidth, i.e, 

\begin{equation}
\nu_i = (1 + \beta\,U_i)\,\nu_\mathrm{m} ,
\label{RandomEq}
\end{equation}

\noindent where $U_i$ is a random real number between 0 and 1, following a uniform statistical 
distribution. We show an example of this distribution of sub-bands in Fig. \ref{DistriFig}(b). 
Different sets of $U_i$ may translate into different fringe dynamic ranges and 
precisions in the estimate of ionospheric dispersion (usually, a higher precision in the estimate
of the atmospheric dispersion translates into a lower dynamic range of the fringes). We 
estimated the quantities $\sigma(K)$, $D$, and $\tau_\mathrm{amb}$ as the averages of those computed from 
100 different sets of $U_i$. Therefore, the results reported in Sect. \ref{IV} for the random 
distribution correspond to an averaged behavior of this distribution. 

\subsection{Power-law distribution}
\label{PowLawSect}

The ionosphere introduces larger phase drifts at lower frequencies. Therefore, it is plausible
that a distribution that samples better the region of lower frequencies will give more precise 
estimates of the ionospheric dispersion, since the phase drifts will be better sampled in the 
region of the spectrum where the ionospheric effects are larger. A natural distribution to 
obtain this kind of sampling is setting the density of sub-bands proportional to a power law 
of frequency, $\nu^{\alpha}$, where $\alpha$ is a given (negative) constant, i.e, 

$$ \int_{\nu_\mathrm{m}}^{\nu_i}{\nu^{\alpha}d\nu} = 
\frac{i-1}{N-1}\int_{\nu_\mathrm{m}}^{\nu_\mathrm{M}}{\nu^{\alpha}d\nu} . $$

Therefore,

\begin{equation}
\nu_i = \left( (\nu_\mathrm{M}^{\alpha+1} - \nu_\mathrm{m}^{\alpha+1})\frac{i-1}{N-1} + \nu_\mathrm{m}^{\alpha+1} \right)^{1/(\alpha+1)} .
\label{PowerEq}
\end{equation}

For the special case of $\alpha = -1$ we have instead

\begin{equation}
\nu_i = \exp{\left( (\ln{\nu_\mathrm{M}} - \ln{\nu_\mathrm{m}})\frac{i-1}{N-1} + \ln{\nu_\mathrm{m}} \right)} .
\label{AlphaMinus1}
\end{equation}

\begin{figure}
\centering
\includegraphics[width=9cm]{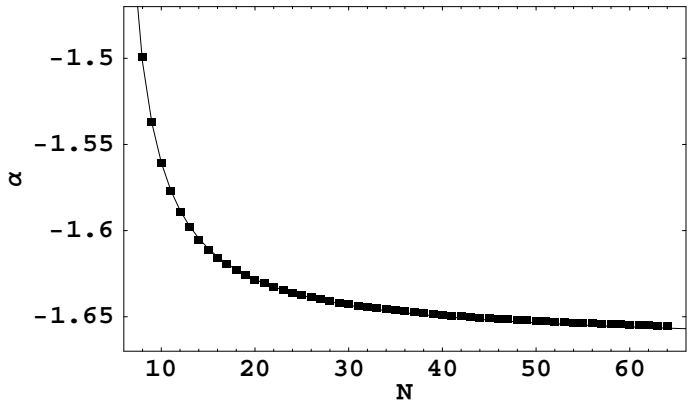}
\caption{Boxes, optimum values of $\alpha$ (i.e., those that minimize the formal uncertainty 
in the estimate of the ionospheric dispersion) as a function of the number of sub-bands. Line,
model corresponding to Eq. \ref{AlphaNEq}.}
\label{AlphaN}
\end{figure}

Equation \ref{PowerEq} becomes Eq. \ref{HomogEq} for $\alpha = 0$.
It can be shown that the standard deviation of $\nu_i^{-1}$ (i.e., the precision in the estimate 
of the ionospheric dispersion) is maximum for $\alpha = -5/3$ in the case of a large number of sub-bands. 
We show in Fig. \ref{AlphaN} the optimum $\alpha$ as a function of the number of sub-bands, $N$. These 
values of $\alpha$ were computed by finding numerically the minimum of $\sigma(K)$ (using Eq. \ref{SigK})
as a function of $\alpha$. The values of $\alpha$ shown in Fig. \ref{AlphaN} can be estimated using the 
model (which is also shown in the figure)

\begin{equation}
\alpha = \frac{0.6}{N - 4.35} - \frac{5}{3} ,
\label{AlphaNEq}
\end{equation}

\noindent which tends to $-5/3$ for large $N$. These are the values of $\alpha$ that were
used to obtain the results reported in the next section. We show an example of this power-law distribution
in Fig. \ref{DistriFig}(c).

\subsection{Golomb rulers}

A Golomb ruler is a set of $n_i$ integers such that the set of differences, $d_{ij} = n_i - n_j$, has
no repeated elements (see, e.g., Atkinson, Santoro \& Urrutia \cite{GolRef}). It is intuitive that Golomb 
rulers are a good choice to maximize the 
dynamic range of the fringes, since all pairs of sub-bands are separated incoherently one respect to 
the other. Therefore, the sidelobes of the Fourier transform of the bandpass are minimum. The improvement
in the fringe dynamic range when the sub-bands are distributed according to Golomb rulers has been 
previously reported for the case of 8 sub-bands (Mioduszewski \& Kogan \cite{Miodus}). Here we generalize 
the study to different 
number of sub-bands and also analyze the impact of this kind of distribution in the precision of the 
estimate of the ionospheric dispersion. The central frequencies, $\nu_i$, of the sub-bands are computed
using the equation

\begin{equation}
\nu_i = \left(1 + \beta\frac{n_i}{n_N} \right)\,\nu_\mathrm{m} ,
\label{GolombEq}
\end{equation}

\noindent where $n_i$ is the $i$th element of a Golomb ruler of $N$ elements (by convention, $n_1 = 0$).
The Golomb rulers for $N < 24$ were taken from the OGR project at \textrm{http://distributed.net/ogr}, 
and the others from Atkinson, Santoro \& Urrutia (\cite{GolRef}). We show an example of this 
distribution in Fig. \ref{DistriFig}(d).

\section{Results}
\label{IV}

Figure \ref{DistriFig} shows that the random and Golomb-rulers distributions tend to poorly 
sample some regions of the spectrum and oversample others. On the contrary, the constant and the 
power-law distributions sample the bandwidth in a more homogeneous way. A more homogeneous sampling of 
the spectrum is preferable to obtain information from as many regions of the bandwidth as possible. 
Moreover, a more homogeneous sampling makes easier the the connection of the phases between 
the sub-bands, since an unsampled wide lag in the spectrum could contain a number of 
$2\pi$ phase cycles that could introduce biases in the data analysis. To 
better understand this 
statement, let us consider, for example, a non-dispersive delay, $\tau$, added to the fringe. 
The differential phase between sub-bands $i$ and $j$, due to $\tau$, would be 

$$\Delta\phi_{ij} = \phi_j - \phi_i = 2\pi\tau(\nu_j - \nu_i)$$

For larger values of $\nu_j - \nu_i$ (i.e., for wider lags between sub-bands), the 
differential phase between sub-bands is larger. Therefore, the probability of the differential 
phase to be larger than $2\pi$ is higher for wider lags between sub-bands.
 
From this point of view, the uniform and/or the power-law distributions would be the best 
frequency configurations for the interferometer. However, we must also 
take into account the dynamic range of the fringes and the quality in the estimate of 
the atmospheric dispersion, which are analyzed in the following subsections.

\subsection{Ionospheric dispersion}

\begin{figure}
\centering
\includegraphics[width=9cm]{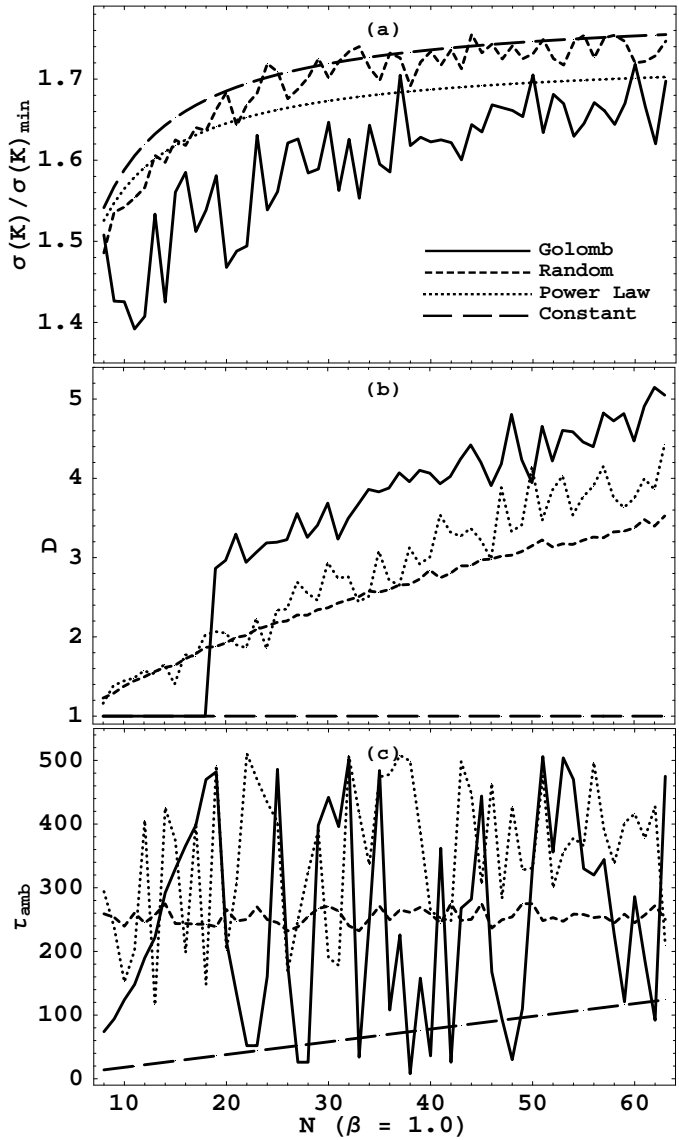}
\caption{(a) Formal uncertainty in the estimate of the ionospheric 
dispersion, in units of the minimum possible uncertainty, (b) dynamic range of the 
fringes, and (c) delay ambiguity (i.e., distance between the fringe peak and the 
closest sidelobe) in units of the Nyquist time resolution. All these quantities are 
shown as a function of the number of sub-bands for $\beta = 1.0$ (see Eq. 
\ref{BetaEq}).}
\label{NFig}
\end{figure}

In Fig. \ref{NFig}(a), we show the uncertainty in the estimate of the ionospheric dispersion, 
$\sigma(K)$, in units of the minimum possible uncertainty (i.e., $\sigma(K)_{min}$, computed from 
Eqs. \ref{SigK} and \ref{minSig}), as a function of the number of sub-bands, for a total bandwidth 
of $\beta = 1.0$ (see Eq. \ref{BetaEq}). The four different distributions are shown. For all 
distributions, the uncertainty (relative to the minimum possible one) increases with the number of 
sub-bands. It can be seen that the distribution based on Golomb rulers give the most precise 
estimates of the ionospheric dispersion, followed by the power-law distribution (with an uncertainty 
$\sim 3$\% larger, depending on the number of sub-bands), and the random and uniform distributions 
(with uncertainties $\sim 6$\% larger, also depending on the number of sub-bands). 

\begin{figure}
\centering
\includegraphics[width=9cm]{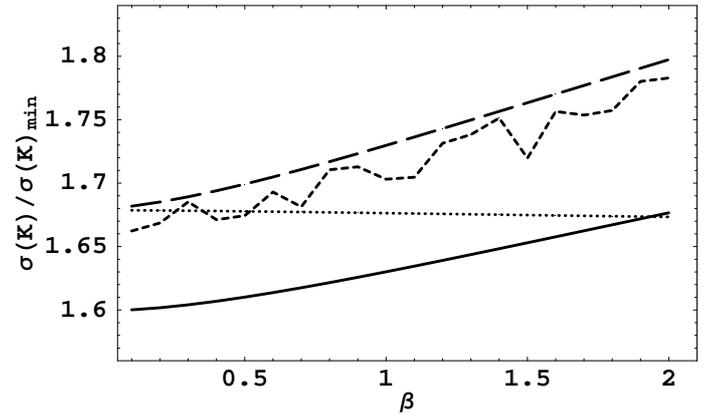}
\caption{Formal uncertainty in the estimate of the ionospheric 
dispersion, in units of the minimum possible uncertainty, as 
a function of $\beta$ (see Eq. \ref{BetaEq}) for the case of 32 sub-bands.}
\label{BetaFig}
\end{figure}

In Fig. \ref{BetaFig}, we show the uncertainty in the estimate of the ionospheric dispersion, in 
units of the minimum possible uncertainty, as a function of the bandwidth (i.e., $\beta$, see Eq. 
\ref{BetaEq}) using a total of 32 sub-bands to cover the bandwidth. Golomb rulers yield again the 
most precise estimates of the ionospheric dispersion, although the uncertainty increases as the 
bandwidth increases. On the contrary, the power-law distribution keeps the uncertainty roughly 
constant as a function of the bandwidth. Both the constant and random distributions also increase 
the uncertainty in the atmospheric dispersion as the bandwidth increases, being this uncertainty 
$\sim$5\% larger than that obtained with the Golomb rulers. 

We conclude that the power-law distribution and that based on Golomb rulers 
give higher precisions in the estimate of the ionospheric dispersion.
Although the difference between uncertainties from all the distributions is not so large
(lower than 10\% in all cases), its optimization may be important to obtain high-contrast 
images.

\subsection{Fringe dynamic range and delay ambiguity}

In Fig. \ref{NFig}(b), we show the dynamic range of the fringes as a function of the 
number of sub-bands for a total bandwidth of $\beta = 1.0$ (see Eq. \ref{BetaEq}).
We notice, however, that the dynamic range of the fringes is independent of $\beta$,
since a change in $\beta$ is equivalent to a change in the delay scaling of the fringes 
(regardless of a phase factor that depends on $\nu_\mathrm{m}$).
Two regions in the sub-band space can be readily seen. 

For $N < 20$, the random and 
the power-law distributions give higher dynamic ranges. Suprisingly, for these 
values of $N$, the distribution based on Golomb rulers give dynamic ranges $\sim 1$.
Why? Golomb rulers are sets of integer numbers. Therefore, the Fourier transforms of 
these sub-band distributions are periodic. If the delay window is larger than the 
period of the Fourier transform, there will be more than one peak in the fringe. The
fringe period depends on each ruler and increases with the number of channels.
For $N<20$, the fringe period is shorter than our delay window (1024 times the 
Nyquist time resolution, i.e., 512 channels in each direction of the delay) so 
there is more than one peak in the fringe. For $N>20$, 
the fringe period is larger and only one peak remains in the delay window. 
For $N > 20$, Golomb rulers give the highest dynamic ranges (around $20-30$\% higher 
than those based on the random and 
power-law distributions). In all cases, the uniform distribution gives very poor
dynamic ranges, $\sim 1$, as it is indeed expected, since the Fourier transform of 
the bandpass is a periodic function with a very short period.
For the case of the delay ambiguity, strong changes are seen as a function of the 
number of sub-bands for the Golomb rulers (the ambiguity ranges between $\sim20$ and 
512 channels) and the power-law distribution (the 
ambiguity ranges between $\sim 150$ and 512 channels, although the lower limit 
increases with $N$). These changes in the delay ambiguity are due to several sidelobes 
with similar peak values. Changing $N$ also changes the relative height of the sidelobe 
peaks. As a consequence, for different values of $N$, different sidelobes are selected 
as the highest and, therefore, very different delay ambiguities are obtained. On the 
contrary, the random distribution has a delay ambiguity of $\sim 250$ channels 
for all values of $N$ (we notice, however, that, for this distribution, the figure shows 
the average of 100 different fringes). The uniform distribution, as expected, has a very 
small delay ambiguity (lower than 100 channels). This ambiguity increases with $N$, also 
as expected, since the spacing of sub-bands (which is shorter for larger $N$) is 
inversely proportional to the period of the fringe.

A first conclusion is that the uniform distribution is not a good choice from the 
point of view of the quality in the estimate of the group delay. The Golomb rulers are 
a good choice when the number of sub-bands is large ($N > 20$, although this number 
decreases if the width of the delay window decreases). 
The power-law distribution is, in general, a good choice for all $N$. It gives the 
best compromise between homogeneity in the sampling of the bandwidth, precision in the 
estimate of the ionsopheric dispersive effects, dynamic range of the correlated fringes, 
and ambiguity of the group delay. Therefore, this would be the preferable sub-band 
distribution to use in low-frequency (wide-band) interferometric observations.

Nevertheless, these conclusions are based on a number of sub-bands up to 64. If the number of 
sub-bands is large (say, $N=512$) there are no available Golomb rulers to work
with, but we can still compare the results obtained from the uniform, random, and 
power-law distributions.

Setting $N=512$, the dynamic range of the fringes is similar for the three distributions, if we use
a delay window of 1024 Nyquist channels (using 512 sub-bands, the period of the fringe corresponding 
to the uniform distribution is longer than the delay window). However, the power-law distribution 
still gives lower formal uncertainties in the estimate of the ionospheric dispersion 
(around 10\% lower than the other distributions for $\beta = 2$ and 4\% for $\beta=1$).
Therefore, the power-law distribution is still the best choice with a number of sub-bands as 
large as 512.

\subsection{Source spectral index}

Wide-band observations allow to precisely determine the spectral
indices and spectral curvatures of radio sources. The different distributions of 
sub-bands may also affect the achievable precision in the 
estimate of the spectral properties of the radio sources. For the case of the 
spectral index, $\gamma$ (being the flux density, $S \propto \nu^{\gamma}$), the 
formal uncertainty, $\sigma(\gamma)$ depends on the distribution $\nu_i$ in the 
form

\begin{equation}
\sigma(\gamma) = \sigma \left(\log{\nu_i} \right)^{-1}.
\label{SigGamma}
\end{equation}

\begin{figure}
\centering
\includegraphics[width=9cm]{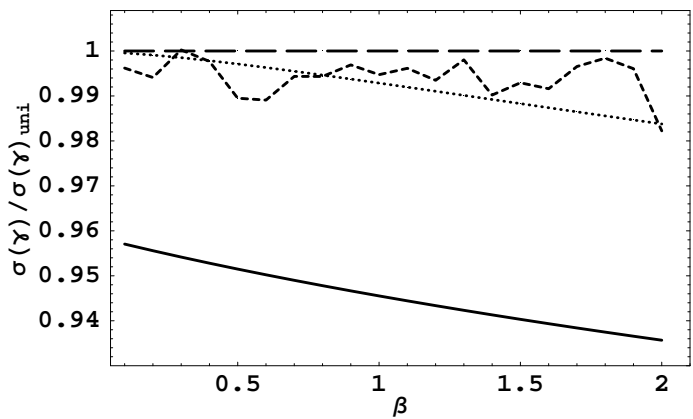}
\caption{Formal uncertainty in the estimate of the source spectral index, $\gamma$, in 
units of that of the uniform distribution ($\sigma(\gamma)_{\mathrm{uni}}$), as
a function of the bandwidth, $\beta$ (see Eq. \ref{BetaEq}), for the case of 64 
sub-bands.}
\label{GammaFig}
\end{figure}

We show in Fig. \ref{GammaFig} the formal uncertainty in the estimate of $\gamma$, 
in units of that of the uniform distribution, $\sigma(\gamma)_{\mathrm{uni}}$, as a function
of the bandwidth, $\beta$, for the case of 64 sub-bands. It can be seen in the figure 
that the uniform, random, and power-law distributions give similar precisions in the 
estimate of the spectral index (although the precision for the power-law distribution 
slightly increases with the bandwidth). Surprisingly, the Golomb-ruler distribution 
gives a precision $\sim5$\% higher than the other distributions for all bandwidths.

The formal uncertainties of the uniform, random, and power-law distributions for 
a large number of sub-bands ($N=512$) are similar to those shown in Fig. \ref{GammaFig}. 
Therefore, for wide-band observations using many sub-bands (i.e. where no Golomb rulers 
are available), the use of the power-law 
distribution is the best choice, allowing for a $1-2$\% higher precision in the estimate 
of the spectral index of the sources. With a smaller number of sub-bands, the use of 
Golomb-ruler distributions would improve the precision in the estimate of $\gamma$ by
$\sim5$\%.

\subsection{Other contributions to the optimum power-law distribution}

Other contributions to the visibility phases (either due to the electron plasma of the ionosphere 
or to chromaticity in the structure of the observed sources), as well as the contribution
of the galactic radiation to the visibility noise, have not been considered in 
the previous sections. In this section, we study how these contributions may affect the 
optimum sampling of the ionospheric dispersion using the power-law distribution of sub-bands. 

\subsubsection{Plasma frequency of ionospheric electrons}
\label{PlasmaFreq}

Equation \ref{IonPhi} holds in the region of frequencies much higher than the plasma frequency,
$\nu_\mathrm{p}$, of the ionospheric electrons. The electron density in the ionosphere takes values around 
$10^4 - 10^6$\,e$^{-}$\,cm$^{-3}$. This translates into a plasma frequency in the range 
$\sim 1 - 10$\,MHz (e.g., Pacholczyk \cite{Pacholczyk1970}, Eq. 2.72). From the refraction index 
of a plasma (e.g., Pacholczyk \cite{Pacholczyk1970}, Eqs. 2.77 and 2.78), the phase drift in the 
case of $\nu_i \sim \nu_\mathrm{p}$ is

\begin{equation}
\Delta \phi_i = K' \nu_i \left( \sqrt{1 + \frac{\nu_\mathrm{p}^2}{\nu_i(\nu_i \pm \nu_\mathrm{B})}} - 1 \right) ,
\label{IonPhi2}
\end{equation}

\noindent where $\nu_\mathrm{B}$ is the Larmor electron frequency (for the Earth magnetic field it 
takes the value $\sim 1$\,MHz) and the $\pm$ signs correspond to the two possible circular polarizations
of the radiation. We notice that the effect of $\nu_\mathrm{B}$ in the computations reported in this section 
is negligible. $K'$ is related to the TEC above each element of the baseline in the line-of-sight 
to the source. 

\begin{figure*}
\centering
\includegraphics[width=18cm]{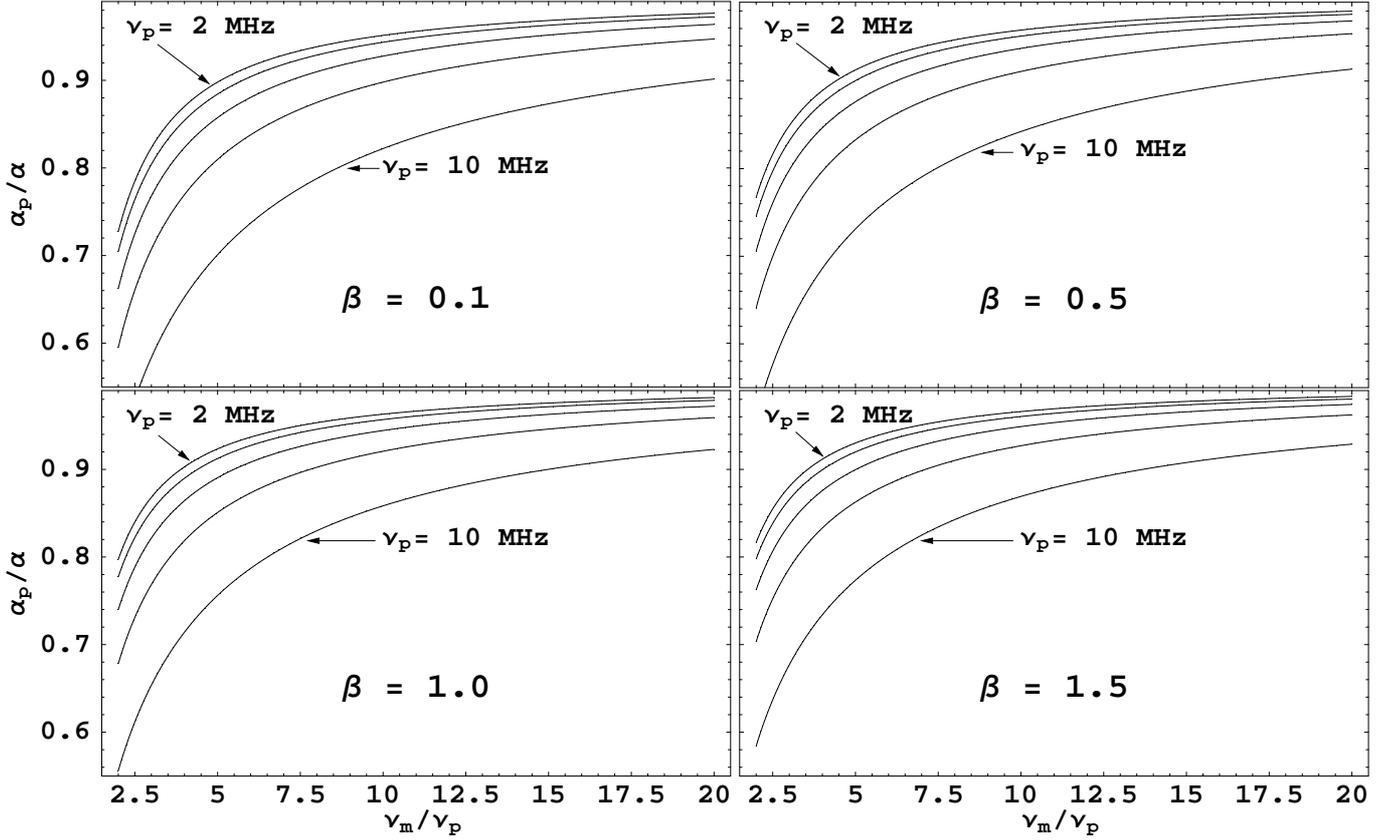}
\caption{Optimum values of $\alpha_\mathrm{p}$, in units of $\alpha$, as a function of the minimum 
observing frequency, $\nu_\mathrm{m}$, in units of the plasma frequency, $\nu_\mathrm{p}$. In each figure, 
the values are computed for five different values of $\nu_\mathrm{p}$ (2, 4, 6, 8, and 10\,MHz). Each 
figure corresponds to a different bandwidth, $\beta$ (0.1, 0.5, 1.0, and 1.5).}
\label{PlasmaFig}
\end{figure*}

We computed the optimum values of $\alpha$ (i.e., the exponent of the power-law distribution of sub-bands) 
that optimize the sampling of the ionospheric phase drifts (i.e., minimize the formal uncertainty in 
the estimate of $K'$) for observing frequencies close to $\nu_\mathrm{p}$. We call these values 
$\alpha_\mathrm{p}$, to distinguish them from the values without the effect of the plasma frequency (i.e., 
$\alpha$, which are shown in Fig. \ref{AlphaN} and given in Eq. \ref{AlphaNEq}). In 
Fig. \ref{PlasmaFig} we show the ratios $\alpha_p/\alpha$ as a function of the minimum 
observing frequency, $\nu_\mathrm{m}$, in units of the plasma frequency, $\nu_\mathrm{p}$.
For instance, using a bandwidth of $\beta = 1.0$ (i.e., $\nu_\mathrm{M} = 2\nu_\mathrm{m}$) a plasma frequency 
$\nu_\mathrm{p}=10$\,MHz, and a minimum frequency $\nu_\mathrm{m} = 100$\,MHz (i.e, 10 times the plasma frequency) 
results in values of $\alpha_p$ equal to 0.86 times those of $\alpha$. For a large number of sub-bands
(i.e., for $\alpha = -5/3$) this results in $\alpha_\mathrm{p} = -1.43$.

We notice that the exponent $\alpha_p$ approaches zero as the minimum frequency approaches to the plasma
frequency. Therefore, the power-law distribution approaches the constant distribution at very low 
frequencies. Increasing the bandwidth tends to compensate a bit the decrease in the absolute 
value of $\alpha_p$ (i.e., the ratios $\alpha_p/\alpha$ increase when $\beta$ increases) although 
this effect is small, as it can be appreciated in Fig. \ref{PlasmaFig}.

\subsubsection{Frequency-dependent source structure}

A source structure which is independent of the observing frequency is a strong assumption over 
the broad frequency ranges considered in the previous sections. The contribution of a possible 
source chromaticity in the visibility phases can be divided in two parts. One is related to the 
source being intrinsically different at different frequencies (this contribution might be especially 
important for extended sources) and the other one is related to the position of the source (or 
that of its brightest feature) being different at different frequencies (as it is the case, for 
instance, of a self-absorbed core-jet structure). 

On the one hand, the contribution
of the source structure to the visibility phases can be determined from the image 
of the source at each frequency. This image depends on the visibility calibration, but can also 
be used to refine such calibration. Therefore, it should be possible, in principle, to decouple the source 
structure (which introduces baseline-dependent phases) from the ionospheric dispersion (which introduces 
antenna-dependent phases), with the help of iterative and elaborated imaging-calibration algorithms. The 
details of these algorithms and their impact in the precision of the estimated ionospheric contribution, as 
it is decoupled from the source-structure contribution after the imaging, is beyond the scope of this paper.

On the other hand, the contribution of a frequency-dependent source position on the visibility phases can be
studied if some assumptions are considered. Porcas (\cite{Porcas}) reported on the effects of source chromaticity 
in the source position estimates through VLBI astrometry, performed using either phase delays or group delays, for 
the case of a core-jet structure following the model of Blandford \& K\"onigl (\cite{Blandford}). The 
contribution of the chromatic core-shift to the interferometric phase is

\begin{equation}
\Delta\phi_{\mathrm{str}} = K_s \,\nu^{1-\delta},
\label{Porcas1}
\end{equation}

\noindent where $K_s$ depends on the physical conditions in the jet and the angle of the 
projected baseline with respect to that of the jet. The parameter $\delta$ also depends on the physical conditions 
in the jet and may take values between 0 and $\sim 2$. If we combine Eqs. \ref{IonPhi} and \ref{Porcas1}, we obtain 
the total drift of the visibility phases through the sub-bands when both effects, ionospheric dispersion and 
source chromaticity, are taken into account:

\begin{equation}
\Delta\phi_i = (K + K_s \,\nu_i^{2-\delta})\,\nu^{-1}.
\label{Porcas2}
\end{equation}

Therefore, the effect of a frequency-dependent source position is equivalent to the 
addition of an extra term coupled to the parameter $K$ of the ionospheric dispersion. For $\delta=0$ (i.e., no 
core-shift), $K_s$ translates into a contribution to $\phi$ equal to a constant (i.e., non-dispersive) group 
delay, with the phase proportional to the frequency. This group delay is equivalent to a shift of the 
source, which is the same at all frequencies. This shift can be easily fixed in the calibration if the source 
position is known. For $\delta=1$, the effect of the source chromaticity on $\phi$ is just adding a constant, so 
it does not affect $\sigma(K)$ (see Eq. \ref{SigK}). Therefore, the sub-band distribution for the optimum 
estimate of ionospheric dispersion will be the same as with no chromatic effect. However, for the (physically 
unrealistic) case $\delta=2$, there is a complete coupling between the ionospheric dispersion, $K$, and the source 
chromaticity, $K_s$, so it is especially difficult to calibrate the ionospheric delay using these sources, 
regardless of the distribution of sub-bands used. Real sources may have values of $\delta$ falling between the values 
here analyzed (0, 1, and 2). It is thus expected to find intermediate cases in which the effect of the 
frequency-dependent source position will be a combination of either a non-dispersive group delay, no effect in the 
estimate of ionospheric dispersion, and a complete coupling between that estimate and the source position. 
Additionally, all the core-jet sources detected in a given FoV may have different values of $\delta$, so different 
calibration issues will appear in the same image depending on the coordinates of each source and the values of 
$\delta$.  

In any case, we notice that $K_s$ depends on the direction of the projected baseline relative to that of the jet, so 
it is a baseline-dependent quantity. However, the ionospheric contribution, $K$, is antenna-dependent. This different 
behavior of $K_s$ and $K$, depending on the pair of stations selected, should allow for a robust decoupling of $K_s$ 
from $K$, provided the number of observing stations is large enough. Therefore, any chromaticity in the 
source structure and/or position should not affect the results reported in this paper, as long as the baseline-dependent
chromatic effects are decoupled from the antenna-dependent ionospheric contribution using the appropriate calibration
algorithms.

\subsubsection{Radiation from the Galaxy}

For frequencies below $\sim400$\,MHz, the sky brightness temperature is dominated by the Galactic radiation, 
which depends strongly on the observing frequency ($T_{\mathrm{sky}} \propto \nu^{\gamma}$ with 
$\gamma \sim 2.5$). It means that in LOFAR wide-band observations, the noise in the low-frequency sub-bands 
will be higher than that in the high-frequency sub-bands. In the cases of observations 
dominated by the radiation from the Galaxy, Eq. \ref{SigK} must be adapted to take into account the different 
uncertainties in each sub-band. 

Thermal noise from the sky brightness temperature translates into a gaussian-like noise in the real and 
imaginary parts of the visibilities, with a value of $\sigma$ proportional to the equivalent flux density of the 
system, which is in turn proportional to the total (i.e., receivers plus source) temperature (e.g., Thomson, 
Moran, \& Swenson \cite{TMS91}). If the observed sources are strong, the noise in the amplitudes and phases can 
also be approximated as being gaussian-like, with a $\sigma$ proportional to that of the real and imaginary parts of 
the visibilities. If we take this approximation into account and assume that the galactic radiation dominates 
the system temperature (i.e., $T_{\mathrm{sys}} \sim T_{\mathrm{sky}}$), then the uncertainty in the visibility
phase of the $i$th sub-band is $\sigma_i \propto \nu_i^{-5}$ and Eq. \ref{SigK} becomes

\begin{equation}
\sigma(K) \propto
\frac{\sum{\nu_i^{5}}}{\sum{\nu_i^{3}}\sum{\nu_i^{5}} - \left(\sum{\nu_i^{4}}\right)^2}.
\label{GalNoise}
\end{equation}

\begin{figure}
\centering
\includegraphics[width=9cm]{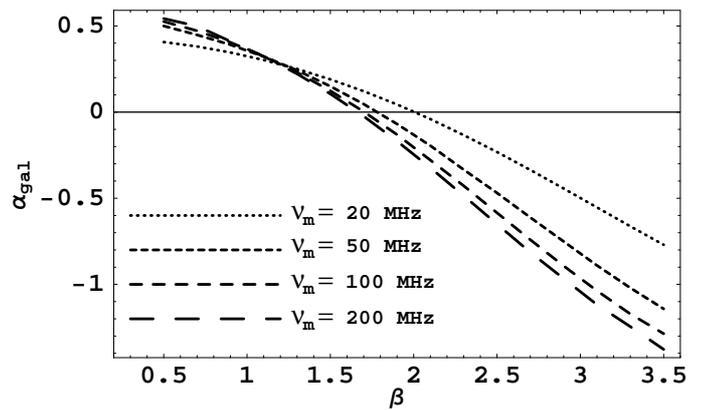}
\caption{Optimum value of $\alpha$ that minimizes the $\sigma(K)$ given in Eq. \ref{GalNoise} for different 
values of the minimum frequency, $\nu_\mathrm{m}$. These values of $\alpha_{\mathrm{gal}}$ have been computed assuming that 
the Galaxy radiation dominates the noise of the visibilities. A line at 
$\alpha_{\mathrm{gal}} = 0$ is also shown for clarity.}
\label{GalFig}
\end{figure}

The values of $\alpha$ in Eq. \ref{PowerEq} that minimize the $\sigma(K)$ given by Eq. \ref{GalNoise}
(we call these values $\alpha_{\mathrm{gal}}$) are shown in Fig. \ref{GalFig} as a function of the 
normalized bandwidth, $\beta$. For small values of $\beta$, we find that $\alpha_{\mathrm{gal}}$ is positive.
It means that the noise in the phases at the lower frequencies is such high, that a better sampling of the 
high-frequency region of the band yields more precise estimates of the ionospheric dispersion. However, this 
effect is less important for wider bands and/or higher minimum frequencies, as it can be seen in Fig. \ref{GalFig}. 
If the band is wide enough (for instance, $\beta > 1.7$ for a minimum frequency of 200\,MHz) the optimum value 
$\alpha_{\mathrm{gal}}$ is negative. The $20-90$\,MHz band of LOFAR corresponds to a value of $\beta$ as large 
as $3.5$, for which we find $\alpha_{\mathrm{gal}} = -0.8$ (setting $\nu_\mathrm{m} = 20$\,MHz).

\section{Summary}
\label{V}

We studied the achievable precision in the estimate of the ionospheric dispersion, the 
ambiguity of the group delay, the dynamic range of the correlated fringes, and the 
precision in the estimate of the source spectral index in 
low-frequency and wide-band interferometric observations for four different distributions 
of the sub-bands through the total bandwidth of the detectors: constant spacing between 
sub-bands, random spacings, spacings based on a power-law distribution, and spacings based 
on Golomb rulers.

For a large number of sub-bands, spacings based on Golomb rulers give the most precise 
estimates of dispersive effects and the highest dynamic ranges of the fringes. Spacings 
based on the power-law distribution give similar (but slightly worse) results, although 
the results are better than those from the Golomb rulers for a smaller numbers of sub-bands. 
Random distributions of sub-bands result in relatively large dynamic ranges of the 
fringes, but the estimate of dispersive effects through the band is worse. A constant 
spacing of the sub-bands results in very bad dynamic ranges of the fringes, but the estimates 
of dispersive effects have a precision similar to that obtained with the power-law 
distribution.

From all combinations of the number of sub-bands and the total covered bandwidth, the power-law 
distribution (with $\alpha$ given by Eq. \ref{AlphaNEq}) gives the best compromise between 
homogeneity in the sampling of the 
bandwidth, precision in the estimate of the ionsopheric dispersive effects, dynamic
range of the correlated fringes, and ambiguity of the group delay. Therefore, this would 
be the preferable sub-band distribution to use in low-frequency (wide-band) interferometric 
observations. 

Finally, we study how the power-law distribution that optimally samples the 
ionospheric dispersion is affected in the cases of 1) observing frequencies close to the 
plasma frequency of the ionospheric electrons, 2 chromatic effects in the structure 
of the sources, and 3) non-negligible noise coming from the Galaxy radiation.

\begin{acknowledgements}

The author is a fellow of the Alexander von Humboldt Foundation in Germany.
The author is very thankful to Eduardo Ros and the anonymous referee for their useful comments 
and suggestions to improve this paper.
The author also acknowledges the collaboration of Nicol\'as Mart\'i-Dunca 
during the preparation of this paper.

\end{acknowledgements}

\end{document}